\documentclass[journal, onecolumn]{IEEEtran}

\IEEEoverridecommandlockouts                              

\usepackage{epsfig} 
\usepackage{amsmath} 
\usepackage{epstopdf}
\usepackage{amsmath} 
\usepackage{amssymb}  
\usepackage{epstopdf}
\usepackage{algorithm}
\usepackage{algpseudocode}
\usepackage{color}
\usepackage{xcolor}

\usepackage{url}
 \usepackage{subfigure}

\usepackage[colorinlistoftodos]{todonotes}

\newcommand{\splitcell}[2][c]{%
  \begin{tabular}[c]{@{}c@{}}\strut#2\strut\end{tabular}%
}

\title{Sample-based SMPC for tracking control of fixed-wing UAV: multi-scenario mapping
}

\author{Martina Mammarella$^{1}$, Elisa Capello$^{2}$, and Fabrizio Dabbene$^{3}$%
\thanks{*This work was not supported by any organization}%
\thanks{$^{1}$M. Mammarella is with the Department
of Mechanical and Aerospace Engineering, Politecnico di Torino, Corso Duca degli Abruzzi 24, 10129 Torino, Italy,
       {\tt\small martina.mammarella@polito.it}}%
\thanks{$^{2}$E. Capello is with the Department of Mechanical and Aerospace Engineering, Politecnico di Torino and with the CNR-IEIIT, Politecnico di Torino, Corso Duca degli Abruzzi 24, 10129 Torino, Italy,            {\tt\small elisa.capello@polito.it}}%
\thanks{$^{3}$ F. Dabbene is with the CNR--IEIIT, Politecnico di Torino, Corso Duca degli Abruzzi 24, 10129 Torino, Italy,
        {\tt\small f.dabbene@ieiit.cnr.it}}%
}

\begin{document}

\maketitle
\thispagestyle{empty}
\pagestyle{empty}

\begin{abstract}
In this paper, a guidance and tracking control strategy for fixed-wing Unmanned Aerial Vehicle (UAV) autopilots is presented.
The proposed control exploits recent results on sample-based stochastic Model Predictive Control, which allow coping in a computationally efficient way with both parametric uncertainty and additive random noise.
Different application scenarios are discussed, and the implementability of the proposed approach are demonstrated through software-in-the-loop simulations.
The capability of guaranteeing probabilistic robust satisfaction of the constraint specifications represents a key-feature of the proposed scheme, allowing real-time tracking of the designed trajectory with guarantees in terms of maximal deviation with respect to the planned one. The presented simulations show the effectiveness of the proposed control scheme.
\end{abstract}

\section{Introduction} 
\label{sec:intro}

Remotely Piloted Aircraft Systems (RPAS) or Unmanned Aerial Vehicles (UAVs), commonly known as drones, are being widely studied and developed due to their mission flexibility, reconfigurable architecture and cost effectiveness. The list of practical and potential applications of UAVs is wide, well beyond the boundaries of conventional remote sensing scenarios, see e.g.~\cite{valavanis,valavanis_handbook} for a discussion. 

\begin{figure}[h!]
\centering
\subfigure[Landslide area]{\includegraphics[width=0.32\columnwidth]{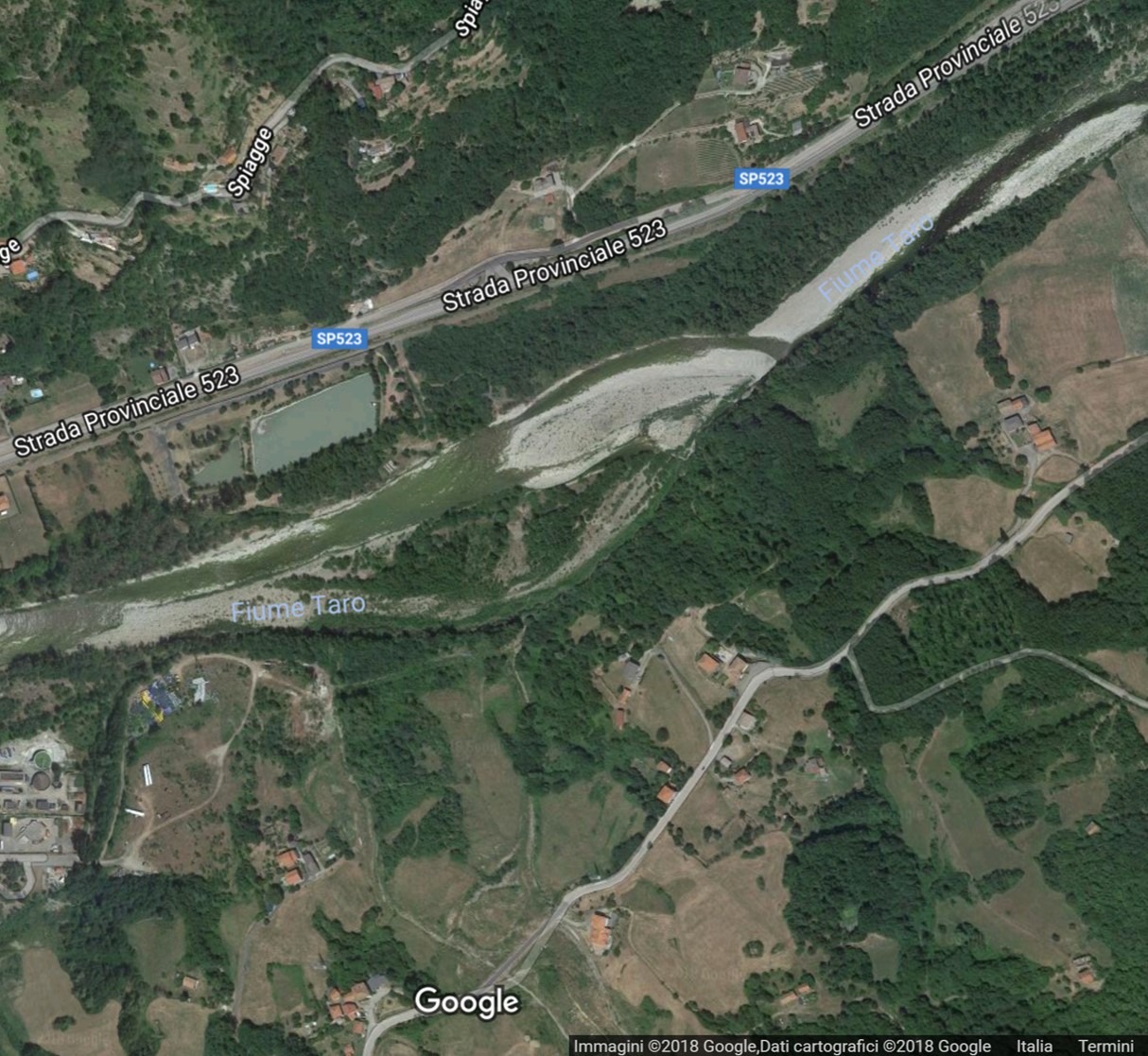}\label{fig:caseA}}
\subfigure[Paddy field]{\includegraphics[width=0.32\columnwidth]{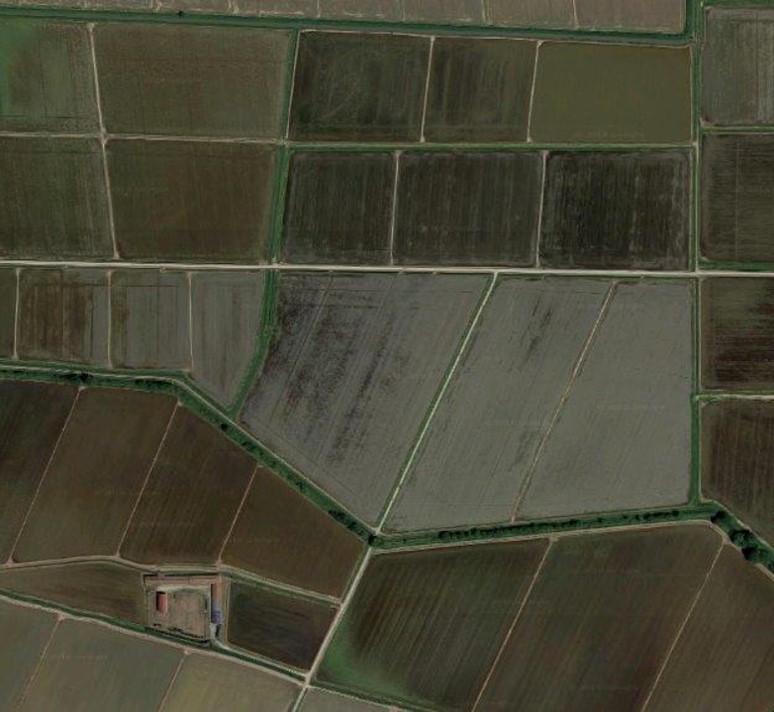}\label{fig:caseB}}
\subfigure[Urban area]{\includegraphics[width=0.32\columnwidth]{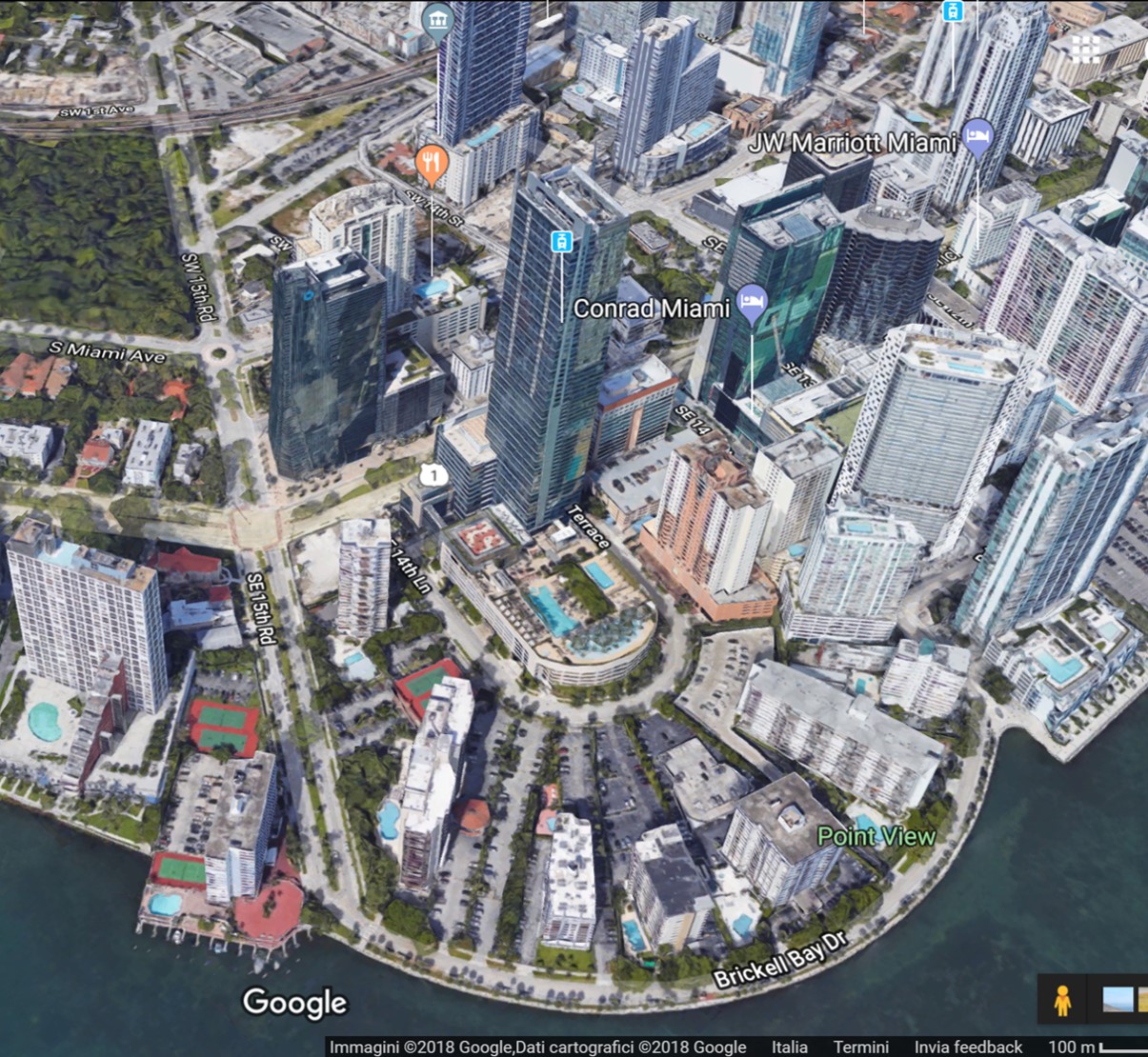}\label{fig:caseC}}
\caption{Overflying area for the three suitable scenarios that could benefit from adopting FW-UAV for mapping purposes.}
\label{fig:miami_sil}
\end{figure}

The majority of studies available in the literature are related to the use of \textit{multi-rotor} UAVs, which are characterized by a high degree of maneuverability and flexibility in configuration. On the other hand, these advantages are paired with some rather serious drawbacks in terms of fuel consumption / maximum mission duration, and relative low speed.
These considerations have recently revived the interest in employing fixed-wing UAVs (FW-UAVs), which are in general able to perform longer missions, cover larger areas, and have lower construction costs.
In this paper, we show how these fields of application are perfectly compatible also with FW-UAVs, investigating three mission scenarios suitable scenarios, which appear to be particularly suited for the use of FW-UAVs but, at the same time, they are rather different in terms of required flight patterns:
\begin{itemize}
\item [(i)] \textbf{rescue and civil protection:} square-patterns, normally covering a pre-assigned sequence of waypoints with selectable levels of persistence (monitoring or ground tracking). In this paper a landslide area is analyzed as an example of rescue monitoring (see Figure \ref{fig:caseA}).
\item [(ii)]  \textbf{precision farming:} grid patterns for territorial coverage, with variable levels of resolution and image overlap, required for an accurate and uniform mapping of the area (crop). An example of the analyzed area (rice field) is in Figure \ref{fig:caseB}. 
\item [(iii)] \textbf{urban monitoring:} complex 3D flight paths, highly connected with the requirements for safety of citizens and collision avoidance, related to buildings, obstacles and other vehicles. In this paper a Downtown Miami's neighborhood is considered
(see Figure \ref{fig:caseC}).
\end{itemize}

For these three scenarios, some common features can be highlighted: (i) high uncertainty and variability, such as the need of producing low-cost FW-UAVs which induces uncertainty in their characteristics; and (ii) safety issues (in particular, the necessity to guarantee mission compliance with guaranteed deviation form the planned trajectory). 

The literature on control design for UAVs is vast, and ranges from proportional derivative integrative (PID) controllers \cite{garcia2012,cap12_aut} to adaptive control \cite{bialy,dydek}. Even if these strategies can guarantee robustness to bounded disturbances, they do not take into account system constraints. From an implementation viewpoint, the class of mini/micro UAVs shares limitations with most computer embedded systems: limited space, limited power resources, increasing computation requirements, complexity of the applications, see e.g.~\cite{pastor}. Moreover, as explained in~\cite{chao}, most of modern autopilots incorporate controller algorithms to meet the always more stringent requirements for the mission accomplishment. The complexity in terms of computational effort of the control algorithms, justifies the trend of most commercial autopilots to  rely on PID controller, which require low computational power.

The combination of guidance and classical MPC schemes for UAV applications, in particular for rotorcraft and multi-rotor systems can be found in literature, see e.g.~\cite{Alexis2016,kamel}. For example, in \cite{Kamel15} a combination of a Linear Quadratic Regulator and Nonlinear MPC (NMPC) is proposed for high-level dynamics. Moreover, in \cite{Kamel17}, a NMPC is proposed including collision avoidance. A fixed-wing application is proposed in \cite{Stastny}, in which high-level dynamics is controlled by a NMPC. The aim of this paper is twofold: (i) real-time implementation of a sample-based Stochastic MPC (SMPC), suitably designed for a customized or open source mini-UAV autopilot, in which the code can be written and implemented by the user, and (ii) a combination of guidance and control algorithms applied to an urban scenario, in which no-aggressive maneuvers are analyzed.

Motivated by the considerations above, the paper develops a framework for FW-UAVs guidance and control, the combination of which can reduce the time of flight and optimize the monitoring of the selected area. 
The proposed approach successfully applies in a real-time framework recent theoretical results obtained by the research team involved in the project: a guidance scheme developed in \cite{cap2014}, and an offline sample based  SMPC approach developed in \cite{cdc2017,TCST}. The use of this latter is important for several reasons: (i) the scheme guarantees probabilistic robust satisfaction of the imposed constraints, at a lower computational cost, and (ii) it is shown that such approach is feasible for medium-high sampling frequencies involved in the application. 
As briefly said before, the effectiveness of the approach for real-time implementation has been validated for a space application on an experimental testbed, as described in \cite{TCST}. In the application here chosen, the goal is more challenging due to the much faster dynamics characterizing the UAV dynamics with respect the spacecraft dynamics, including atmospheric disturbances (i.e. additive noise), model uncertainties (variations on speed $V$ and mass $m$) and platform inaccuracies (variations on the moments of inertia $J$). 
The framework is tested by software-in-the-loop simulations and the main results are presented in Section \ref{sec:results}.

\section{Fixed-wing-UAV and on-board Systems}
\begin{figure}[thpb]
\centering
\subfigure{\includegraphics[width=0.45\columnwidth]{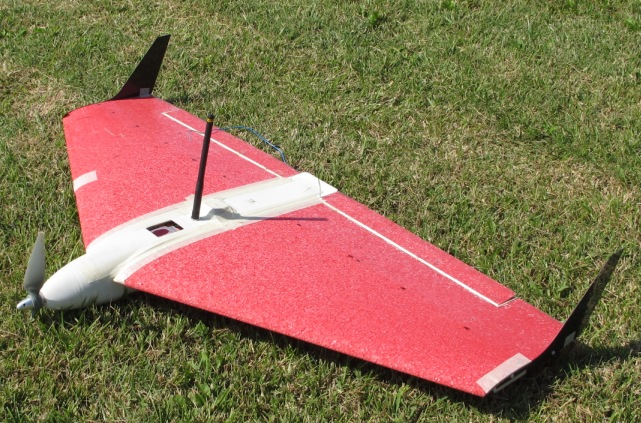}\label{MH850}}
\subfigure{\includegraphics[width=0.45\columnwidth]{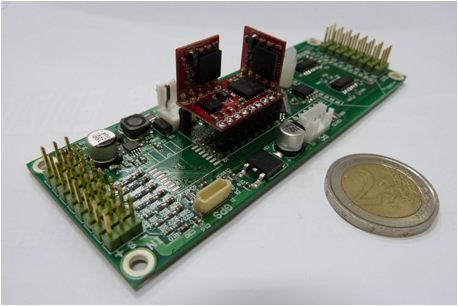}\label{autopilot}}
\caption{ The MH850 mini-UAV (left) and the custom-made autopilot board designed and produced in the Politecnico di Torino DIMEAS (right).}
\end{figure}

The MH850 fixed-wing UAV in Figure \ref{MH850} is one of the platforms of the {\it MicroHawk family} of UAVs \cite{marg}. It was developed for low cost alpine surveillance missions, in which high altitude, low temperature and strong winds represent critical conditions of the mission. This platform has enough specific excess power to climb with non-marginal rates at altitude \cite{marg}. Usually, in piloted-mode, only 45\% of the maximum power is required for level flight (at sea level).
This FW-UAV has an expanded poly-propylene (EPP) wing and a sintered nylon fuselage, with trailing edge elevon (symmetric deflection for elevator $\delta_e$ and antisymmetric for aileron $\delta_a$), and is equipped by an electric brushless motor. A Lithium polymer battery of 5000 mAh with two cells 2S (7.4 V) with a discharge efficiency of $0.875$. With this battery a small camera as a payload, the endurance is about $70$ min at an airspeed of $13.5$ m/s and at an altitude of $100$ m. 
A database including all the aerodynamic derivatives was employed to design the linear and nonlinear aircraft models \cite{marg}.

A complete nonlinear model (as defined in \cite{etk}) is a set of nine equations describing the forces, moments, angles and angular speeds which characterize the flight condition of the aircraft.

These equations of motion are referred to a body reference frame, that is fixed in the aircraft. Classical assumptions of rigid body and flat non-rotating Earth are made. These assumptions are
supported by the application to a Mini-UAV. The state variables in the longitudinal plane are the longitudinal component of the total airspeed $u$, the vertical component of the total airspeed $w$ (and, as a consequence, the angle of attack $\alpha \simeq \frac{w}{V}$), the pitch angle $\theta$ and the pitch rate $q$.
The lateral-directional states are the lateral component of the total airspeed $v$, the roll rate $p$, the yaw rate $r$ and the roll angle $\phi$. 
The aircraft control input is based on trailing edge elevon (symmetric deflection for elevator $\delta_e$ and antisymmetric for aileron $\delta_a$).
The components of the total speed $V$ can be expressed as follows
\begin{eqnarray*}
\begin{aligned}
\dot{u} =   \frac{F_X}{m}+qw-rv+g\sin\vartheta ,\\
\dot{v} =  \frac{F_Y}{m}-pw+ru-g\cos\vartheta\sin\phi,\\
\dot{w} =  \frac{F_Z}{m}+pv-qu-g\cos\vartheta\cos\phi ,
\end{aligned}
\end{eqnarray*}
where $m$ is the aircraft mass, $[u \ v \ w]^T$ are respectively the longitudinal, lateral and vertical components of the speed $V$, $[F_X \ F_Y \ F_Z]^T$ are the forces acting on the three reference axes, $[p \ q \ r]^T$ are the angular speeds along the three axes, $\vartheta$ is the pitch angle and $\phi$ is the roll angle. 

The variation of the angular speeds $[p \ q \ r]^T$ is expressed by the following equations
\begin{eqnarray*}
\begin{aligned}
\dot{p} = \frac{L}{J_x}+ [J_{xz}(\dot{r}+pq)+qr(J_y-J_z)]{J_x},\\
\dot{q} = \frac{M}{J_y}+ \frac{[J_{xz}(r^2-p^2)+pr(J_z-J_x)]}{J_y},\\
\dot{r} = \frac{N}{J_z}+ \frac{[J_{xz}(\dot{p}-pq)+pq(J_x-J_y)]}{J_z} ,
\end{aligned}
\end{eqnarray*}
where $[L \ M \ N]^T$ are the roll, pitch and yaw moments respectively, $J_i$ are the moments of inertia with $i=x ,y ,z ,xz$.

The variation of the Euler angles $[\phi \ \vartheta \ \psi]^T$, which represents the aircraft attitude, is defined by the kinematic equations and it can be obtained integrating the following  equations
\begin{eqnarray*}\label{kinem}
\begin{aligned}
\dot{\phi} = p+q\sin\phi\tan\vartheta+r\cos\phi\tan\vartheta,\\
\dot{\vartheta} = q\cos\phi -r\sin\phi,\\
\dot{\psi} =\frac{q\sin\phi}{\cos\vartheta}+\frac{r\cos\phi}{\cos\vartheta}.
\end{aligned}
\end{eqnarray*}
For the evaluation of the aircraft navigation, the position vector $[x \ y \ h]^T$ is considered 
\begin{eqnarray*}
\begin{aligned}
V_N  =  u\cos\vartheta\cos\psi+v(\sin\phi\sin\vartheta\cos\psi-\cos\phi\sin\psi)+\\
w(\cos\phi\sin\vartheta\cos\psi+\sin\phi\sin\psi),\\
V_E  =  u\cos\vartheta\sin\psi+v(\sin\phi\sin\vartheta\sin\psi+\cos\phi\cos\psi)+\\
+ w(\cos\phi\sin\vartheta\sin\psi-\sin\phi\cos\psi),\\
V_D = u\sin\vartheta+v\cos\vartheta\sin\phi+w\cos\phi\cos\vartheta ,
\end{aligned}
\end{eqnarray*}
where $[V_N \ V_E \ V_D]^T$ are the components of the total airspeed along the three axes in the vehicle-carried vertical reference frame. This frame is centered in the aircraft center of gravity. The axis $X_V$ is directed North, the axis $Y_V$  is directed East and the axis $Z_V$ is directed along the local gravity acceleration vector.

Moreover, for the evaluation of the computational effort and for the evaluation of the battery consumption, the following model of the battery is implemented.

\begin{eqnarray}
\label{battery}
I = I_0 +\frac{dI}{d\tau}\tau\\
V = (\omega-r_1\tau)\frac{V_0}{\omega_0},\\
\end{eqnarray}
with
\begin{eqnarray*}
r_1 = -\frac{\omega_0}{\tau_{ST}}\\
\frac{dI}{d\tau} = \frac{I_{ST}-I_0}{\tau_{ST}},
\end{eqnarray*}
where $I_0$ is related to the lost of current, $I_{ST}$ is the stall current and $\tau_{ST}$ is the stall torque. $V_0$ and $\omega_0$ are the nominal voltage and rpm values. The torque $\tau$ is evaluated starting from the blade element theory \cite{padfield}, using a Clark-YM15 airfoil with Xx mm of diameter. 
The charge level at the time $t+1$ is
\begin{equation*}
E^{t+1}_b=E^t_b-\frac{V^t I^t}{\eta_b}t,
\end{equation*}
with $\eta_b = 0.85$ propeller efficiency and $I^t$ and $V_t$ are the current and the voltage, respectively, evaluated at the time $t$ as in Eq.~(\ref{battery}).

A custom-made autopilot is installed on-board, designed and produced in the Department of Mechanical and Aerospace Engineering (DIMEAS) of Politecnico di Torino \cite{cap13} (see Figure \ref{autopilot}). Main characteristics comprehend an open architecture, the possibility to be reprogrammed in flight, and real-time telemetry. Sensors include GPS, barometric sensor, differential pressure sensor, and three-axis gyros and accelerometers.
The CPU is the ATXMEGA256A3U-3U model with 256Kb flash memory and 16Kb of RAM. A Radiomodem Xbee Pro S1 is used for the communication link between the Ground Control Station (GCS) and the autopilot.

The system can be equipped with different on-board sensors, depending on the defined monitoring mission. For example, for precision farming, a multispectral camera and infrared (IR) cameras, e.g.\ a passive sensor as the RedEdge-M$\textsuperscript\textregistered$ crop sensor from MicaSense (www.micasense.com). This sensor provides, after processing, the Normalized Difference Vegetation Index (NDVI), which is used for the evaluation of the intensity radiation of some bands on the electromagnetic spectrum reflected by canopy (upper portion of plants).
For urban monitoring and risk mapping, a camera and video telemetry can be installed on-board. For example, a 2 mega pixel, 480 TVL camera can be placed on board to be used in real-time for performing surveillance, safety inspections and other video-based operations possible. In a similar way, a video telemetry module transmits video data from FW-UAV to the Ground Control Station (GCS).

\section{Guidance and Control Strategy}
\subsection{Guidance Algorithm}
\label{sec:guidance}
The guidance algorithm adopted in this study is  described in details in \cite{cap2014}, in which some simplifying hypotheses, according to the flash memory limitation of the autopilot micro controller, are taken into account. A given set of
waypoints is considered, with assigned North, East and
altitude coordinates. This set of waypoints includes the
starting point, which is the point where the FW-UAV finishes its
climb and the autonomous flight starts. The starting point and
all the waypoints are assumed to be at the same altitude; thus,
a 2D path is considered. A trajectory smoother, that renders kinematically feasible the assigned trajectory in terms of speed and turn rate
constraints, is implemented.
For the evaluation of the performance of the guidance algorithm for aerial mapping, the cross-track error (CTE) $\epsilon_r$ is calculated, to monitor the FW-UAV position with respect to the reference path \cite{cap2014}. The FW-UAV real position $P_{UAV}$ is considered in terms of East and North coordinates, i.e. $E_{UAV}$ and $N_{UAV}$ respectively, and the segment connecting two waypoints in terms of previous waypoint WP$_n (E_n, N_n)$ and next waypoint WP$_{n+1}(E_{n+1}, N_{n+1})$.
The CTE is then calculated as
\begin{equation}
\label{eq:cte}
\epsilon_r= \frac{\vert E_{UAV}-mN_{UAV}-(E_n-mN_n)}{\sqrt[]{m^2+1}} ,
\end{equation}
with $m=\frac{E_{n+1}-E_n}{N_{n+1}-N_n}$.

A look-ahead or proximity distance is included to define the minimum
distance of the UAV from the next waypoint. When the distance between the aircraft and the next waypoint is less than this pre-defined value, the waypoint is reached and the aircraft can move to the next waypoint.

\begin{figure}[h!]
\centering
\subfigure[Guidance phases]{\includegraphics[width=0.45\columnwidth]{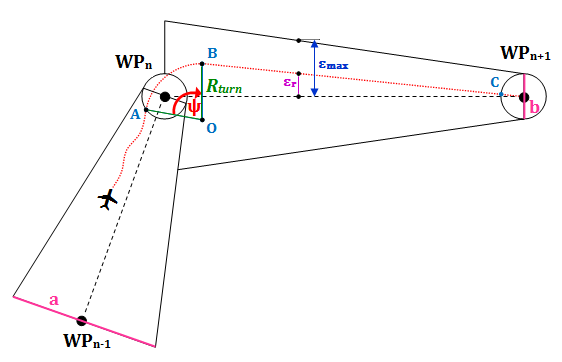}%
\label{fig:guidance1}}
\hfil
\subfigure[CTE and reference distances]{\includegraphics[width=0.45\columnwidth]{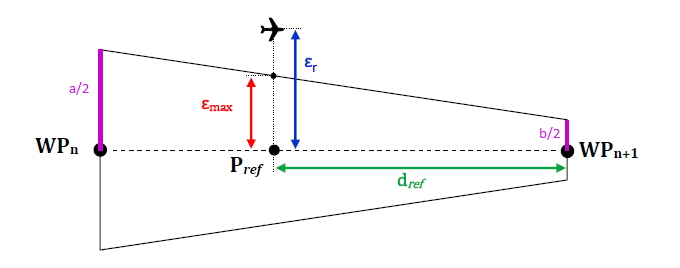}%
\label{fig:guidance2}}
\caption{Guidance algorithm scheme.}
\label{fig:guidance}
\end{figure}


Starting from these hypotheses, the guidance sequence can be divided in three phases, as  described in \cite{cap2014}.
The first phase is the waypoint approach. Assuming that the aircraft is flying with a fixed speed at a defined altitude, this phase is identified by the red dotted line before  point A in Figure \ref{fig:guidance1}.
It is assumed that the waypoint has been reached when the
FW-UAV flies into the imaginary circle centered in the waypoint WP$_n$ with radius equal to the defined proximity distance. In our case the proximity radius is $20$ m, according to the MH850 dynamic constraints (in terms
of minimum turn radius).
In the second phase, the aircraft turns around the waypoint, according with the turn rate constraints. In Figure \ref{fig:guidance1}, this segment is identified by the red dotted curve between  points A and B.
Finally, the last phase is related to the straight flight that starts at the end of the last turn and ends at the beginning of the next turn, represented by the red spotted
line between the points B and C in Figure \ref{fig:guidance1}. In this phase, the CTE is applied (as in \cite{cap2014}).
Moreover, a no correction zone is included, in which the corrections on the heading angle are imposed only when the FW-UAV CTE is larger than an assigned
value (i.e. maximum acceptable CTE). 

\subsection{Tracking-SMPC Algorithm Based on Offline Sampling}
\label{sec:control}
To control the FW-UAV dynamics, we develop upon the approach proposed in \cite{cdc2017}. Our goal is to slightly modify this initial framework in order to obtain a flexible reference tracking algorithm that still shares the same properties. To this end, let us consider a typical servo-problem, where the main objective is to track a reference signal $r_k$. Hence, the desired output, at steady state, shall be $y_k=r_k$. Here, we focus on state-tracking, hence we consider the full-information case, with $C=I$. This setup well suits the application case under consideration, in which we have access to all states of the FW-UAV dynamics.

Formally, we start from the following discrete-time system subject to both random noise and stochastic uncertainty
\begin{align}
\label{eq:sys}
x_{k+1} &= A(q_{k})x_{k}+B(q_{k})u_{k}+w_{k},
\end{align}
with state $x_{k} \in \mathbb{R}^{n}$, control input $u_{k} \in \mathbb{R}^{m}$, output $y_{k} \in \mathbb{R}^{n}$, additive disturbance $w_{k} \in \mathbb{R}^{m_{w}}$, and parametric uncertainty $q_{k}\in \mathbb{R}^{n_{q}}$. We assume that $A(q_{k})$ and $B(q_{k})$, of appropriate dimensions, are functions of the uncertainty $q_k$, which is bounded within the convex polytope $\mathbb{Q}$. These uncertainties are realization at time $k$ of independent and identically distributed (i.i.d.) multivariate real valued random variables $Q_{k}$. Moreover, the disturbance sequence $(w_{k})_{k\in \mathbb{N}_{\geq 0}}$ is assumed to be a realization of iid, zero-mean random variables $(W_{k})_{k\in \mathbb{N}_{\geq 0}}$, with bounded and convex support $\mathbb{W}$.

To achieve tracking, at steady state we shall have $x_{k+1}=x_k=r_k$. Hence, the system dynamics \eqref{eq:sys} can be rewritten in terms of the state deviation $\delta x_k$ with respect to the reference $r_k$, i.e. $\delta x_k=x_k-r_k$, as follows
\begin{equation}
\delta x_{k+1} = A(q_{k})\delta x_{k}+B(q_{k})u_{k}+w_{k}.
\label{eq:sysx_delta}
\end{equation}
The system is subject to $p_x$ individual chance-constraints on each $\alpha$-th state and hard constraints on the \textit{m} inputs
\noindent
\begin{subequations}
\begin{align}
\label{eq:constr_1}\
\!\!\!\mathbb{P}\left\{[H_{x}]_{\alpha}\delta x_{\ell|k}\leq [h_x]_\alpha \right\}\geq1-\epsilon_{\alpha}, &\; \forall \, \ell\in \mathbb{N}_{\geq0},\, \alpha\in \mathbb{N}_{1}^{p_x}\\
\label{eq:constr_2}\
H_{u} u_{\ell|k}\leq h_u, &\; \forall\, \ell\in \mathbb{N}_{>0},
\end{align}
\label{eq:constr}
\end{subequations}
for each predicted $\ell$ step ahead at time $k$ and with $\epsilon_{\alpha}\in (0,1)$. Each chance-constraint represents the maximum admissible deviation of the correspondent state, defined in compliance with the mission under analysis. Following one of the typical approaches in stabilizing MPC, we consider that a suitable terminal set and an asymptotically stabilizing control gain for (\ref{eq:sysx_delta}) exists. Formally, we assume that there exists a terminal set $\mathbb{X}_{T}=\left\{\delta x_k\,|H_{T}\delta x_k\leq h_T\right\}$, which is robustly forward invariant for (\ref{eq:sysx_delta}) under the (given) control law $u_{k}=K\delta x_{k}$. That is, given any $\delta x_{k}\in \mathbb{X}_{T}$, the state and input constraints (\ref{eq:constr}) are satisfied and there exists $P\in \mathbb{R}^{n\times n}$ such that
\begin{equation}
Q+K^{T}RK+\mathbb{E}[A_{cl}(q_k)^{T}PA_{cl}(q_k)]-P\preceq 0
\end{equation}
for all $q_k\in\mathbb{Q}$, with $A_{cl}(q_k)\doteq A(q_k)+B(q_k)K$, and with $Q\in \mathbb{R}^{n\times n}$, $Q\succ 0$, $R\in \mathbb{R}^{m\times m}$, $R\succ 0$.

We then follow a dual-mode prediction scheme typical of robust MPC \cite{kouva}, and consider the design of a parametrized feedback policy
of the form
\begin{equation}
u_{\ell|k}=K\delta x_{\ell|k}+v_{\ell|k},
\label{eq:feedback}
\end{equation}
where for a given $\delta x_{0|k}=\delta x_{k}$, the correction terms $\textbf{v}_k\doteq\left\{v_{\ell|k}\right\}_{\ell\in \mathbb{N}_{0}^{T-1}}$ are determined by the SMPC algorithm as the minimizer of the expected finite-horizon cost
\begin{equation}
\small
J_{T}(\delta x_{k},\textbf{v}_{k})\!\!=\!\!\mathbb{E}\left\{\sum_{\ell=0}^{T-1}(\delta x_{\ell|k}^{T}Q\delta x_{\ell|k}+u_{\ell|k}^{T}Ru_{\ell|k})
+\delta x_{T|k}^{T}P\delta x_{T|k}\!\!\right\},
\label{eq:cost}
\end{equation}
subject to constraints \eqref{eq:constr}.

\subsection{Offline Uncertainty Sampling for SMPC}
In this section, following the approach proposed in \cite{cdc2017,TCST}, we derive \textit{offline} a sample-based inner approximation for the chance constraint \eqref{eq:constr}, obtaining linear constraints on $\delta x_k$, $\textbf{v}_k$ to be adopted for the optimization control problem. In particular, the considered SMPC scheme develops as follows: first, equation (\ref{eq:sys}) is explicitly solved with prestabilizing input (\ref{eq:feedback}) for the predicted states $\delta x_{1|k},\ldots,\delta x_{T|k}$, and predicted inputs $u_{0|k},\ldots,u_{T-1|k}$. Simple algebraic manipulations show that it is possible to derive suitable transfer matrices 
\[
\Phi_{\ell|k}^{0}({q}_{k}), \quad \Phi_{\ell|k}^{v}({q}_{k}),\quad \Phi_{\ell|k}^{w}({q}_{k}) \text{ and } \Gamma_{\ell},
\] 
whose definition can be found in \cite[Appendix A]{cdc2017} for details, such that
\begin{subequations}
\small
\begin{align}
\label{eq:state_new}
\!\!\delta x_{\ell|k}({q}_{k},\textbf{w}_{k})
\!\!&=&\!\!\!\!\!
\Phi_{\ell|k}^{0}({q}_{k})\delta x_{k}+\Phi_{\ell|k}^{v}({q}_{k})\textbf{v}_{k}+\Phi_{\ell|k}^{w}({q}_{k})\textbf{w}_{k}\\
\label{eq:input_new}
\!\!u_{\ell|k}({q}_{k},\textbf{w}_{k})
\!&=&\!\!\!\!\!
K\Phi_{\ell|k}^{0}({q}_{k})\delta x_{k}+(K\Phi_{\ell|k}^{v}({q}_{k})+\Gamma_{\ell})\textbf{v}_{k}\nonumber \\ 
&&+K\Phi_{\ell|k}^{w}({q}_{k})\textbf{w}_{k},
\end{align}
\label{eq:sys_new_1}
\end{subequations}
where $\textbf{w}_k\doteq\left\{w_{\ell|k}\right\}_{\ell\in \mathbb{N}_{0}^{T-1}}$.
In these equations,  both predicted states and inputs are function of the uncertainty $q_k$ and the noise sequence $\textbf{w}_{k}$. Then, we note that, from (\ref{eq:sys_new_1}), the expected value in (\ref{eq:cost}) can be computed offline, leading to a deterministic quadratic cost function of the form $J_{T}(\textbf{z}^T_{k})=\textbf{z}_{k}^T\tilde{S}\textbf{z}_{k}$  in the variable $\textbf{z}_k\doteq[\delta x^T_{k},\textbf{v}^T_{k}]^T$ (see \cite[Appendix A]{cdc2017} for details and definition of $\tilde{S}$).

Second, as described in \cite{cdc2017}, an estimate of the chance constraint sets corresponding to linear constraints \eqref{eq:constr} may be constructed extracting a number of i.i.d. samples $q^{(i)}$ from $Q_k$, which is defined exploiting results from statistical learning theory \cite{vidyasagar}. 
This is obtained by extracting, for $\ell\in\mathbb{N}_0^{T-1}$, $N_{\ell}^x$ i.i.d. samples of the uncertainty and of the noise, according to their respective distribution, obtaining   
the following \textit{tightened} linear constraints
\begin{equation}
H_{x}\delta x_{\ell|k}({q}^{(i_x)},\textbf{w}^{(i_x)})\leq h_x,
\label{eq:EQ1}
\end{equation}
for $i_x\in\mathbb{N}_{1}^{N_{\ell}^x}$ and $\ell\in\mathbb{N}_{0}^{T-1}$.
In a similar way, the hard input constraints and the terminal constraints can be approximated by drawing offline respectively $N_{\ell}^{u}$ and $N_{T}$ samples, obtaining linear constrain
 \begin{equation}
H_{u}u_{\ell|k}({q}^{(i_u)},\textbf{w}^{(i_u)})\leq h_u,\\
\label{eq:EQ2}
\end{equation}
\begin{equation}
H_{T}\delta x_{T|k}({q}^{(i_T)},\textbf{w}^{(i_T)})\leq h_T,\\
\label{eq:EQ3}
\end{equation}
In \cite{TCST} the reader can find details on the construction of the matrices involved.

The above linear constraints constitute a probabilistically guaranteed approximation of the actual constraint observed online. In particular,  it was shown in \cite{matthias1} that if, for given probabilistic parameters $\epsilon_{\alpha}, \epsilon_{\beta}, \epsilon_{\gamma} \in (0,0.14)$, and $\delta \in (0,1)$, the samples are drawn such that 
 $N_\ell^x\ge\tilde{N}(n+\ell m,\epsilon,\delta)$, 
 $N_\ell^{u}\geq \tilde{N}(n+\ell m,\epsilon_{\beta},\delta)$,
 $N_{T}\geq \tilde{N}(n+Tm,\epsilon_{\gamma},\delta)$, with $\tilde{N}(\cdot,\cdot,\cdot)$
given in \cite{TCST}.
The  linear constraints (\ref{eq:EQ1}), (\ref{eq:EQ2}), (\ref{eq:EQ3}), possibly after constraint reduction, can be summarized in the following linear constraint set 
\begin{align}
\mathbb{D}=\left\{
\textbf{z}_k\,\,\,|\,\,\, \tilde{H}\textbf{z}_k\leq \tilde{h}\right\}.
\label{eq:final_constr}
\end{align}
Moreover, as in \cite{cdc2017}, a first step constraint set is added to \eqref{eq:final_constr}, to ensure robust recursive feasibility, given by
\begin{eqnarray}
\nonumber
&\!\!\mathbb{D}_{R}=&\left\{\delta x_{k},\textbf{v}_{k}\,\,\,|\,\,\,H_{\infty}A_{cl}(q_k)\delta x_{k}+H_{\infty}B(q_k)v_{0|k}\leq \right. \\
&& \quad\left. h_{\infty}-H_{\infty}w_{0|k}, \,\,\,j\in \mathbb{N}_{1}^{N_{c}}\right\}\label{eq:first_step_constr}
\end{eqnarray}
with $A(q_k), B(q_k)$ from Assumption 1 and $A_{cl}(q_k)=A(q_k)+ B(q_k)K$. The final set of linear constraints to be employed in online implementation is thus given by the intersection of the sets $\mathbb{D}$ and $\mathbb{D}_{R}$, defined in \eqref{eq:final_constr} and \eqref{eq:first_step_constr} respectively.

The  algorithm, in which a computationally light implementation phase is paired with rather intense offline step is reported next\\

\noindent
\textbf{Offline-sample-based SMPC Algorithm}\\
\noindent\textsc{Offline Step.} {\it Before running the online control algorithm:}
\begin{enumerate}{\it
\item Compute matrix $\tilde S$; draw uncertainty and noise samples
 and construct  sets $\mathbb{D}$ and $\mathbb{D_{R}}$
in \eqref{eq:first_step_constr}.}
\end{enumerate}
\textsc{Online Implementation.}  {\it At each time step $k$:}
\begin{enumerate} {\it 
\item Measure the current state $x_{k}$;
\item Solve the quadratic problem
\[
\label{eq:algo}
\textbf{v}_{k}^{*}  =\arg\min_{\textbf{z}_{k} \in \mathbb{D}\cap \mathbb{D}_{R}} \textbf{z}_{k}^T \tilde{S} \textbf{z}_{k}
\]
\item Apply the control input 
$
u_{k}=K\delta x_{k}+v_{0|k}^{*},
$
where $v_{0|k}^{*}$ is the first control action of the optimal sequence $\textbf{v}_{k}^{*}$.}
\end{enumerate}

\section{Software-In-the-Loop Results}
\label{sec:results}
The SMPC algorithm described in the previous section is used to control a FW-UAV, whose systems-equation descriptions are given in \cite{stevens}.  
In particular, we consider a linear case, in which the longitudinal and lateral-directional motions result to be decoupled, and each of them can be modeled in the standard discrete time-invariant state-space formulation as \eqref{eq:sysx_delta}, where $A,B$ represent the discrete-time state and input matrices respectively, obtained discretizing the corresponding continuous ones derived starting from the equations in  \cite{ICUAS18}.
The SMPC controller is adopted to control
both the longitudinal dynamics, including airspeed longitudinal component $u$, angle of attack $\alpha$, pitch angle $\theta$, pitch rate $q$ and altitude $h$, as well as the lateral-directional dynamics in terms of airspeed lateral component~$v$, roll and yaw rates, $p$ and $r$ respectively, and roll angle $\phi$. 
The states are fully measured and the angle of attack $\alpha$ is evaluated from the Pitot tube sensor (i.e. differential pressure sensor) and from the accelerometers.

The controller provides the control actions in terms of throttle $\Delta T$, and elevator and aileron deflections, $\delta_e$ and $\delta_a$ respectively. Finally, a PID controller is adopted for the heading angle $\psi$, obtaining the reference roll angle $\phi_{ref}$. Moreover, the other reference signals, $u_{ref}$, $h_{ref}$ and $\psi_{ref}$, are provided by the guidance algorithm as a function of the identified waypoints, as described in Section \ref{sec:guidance}. 

For the evaluation of the battery discharge, the stall current is $I_{ST} = 389.50$ A and the stall torque is $\tau_{ST} = 14.9$ Nm. A loss of current of $I_0 = 0.8$ A is considered and $\omega_0 = 7562.8$ rpm is the zero value of the angular speed.

As anticipated before, three different scenarios have been analyzed, each one characterized by a different mission goal, overflying area configuration and path-to-follow: (i) a high-risk landslide area as civil protection scenario; (ii) a paddy-field in the precision farming framework; (iii) a highly populated neighborhood in Downtown Miami for urban monitoring.

For all the analyzed scenarios, the initial conditions have been set as follows: (i) altitude $h_0=100$ m, (ii) airspeed $V_0=13.5$ m/s, (iii) angle of attack $\alpha_0=5.18$ deg, and (iv) ramp angle $\gamma_0=0$ deg. The multi-rate simulator adopted to validate the guidance and control strategy is characterized by two different sample times, as follows: (i) system sample time $0.01$ s; (ii) SMPC sample time $=0.1$ s. For what concern the main SMPC parameters, adopted for all the case studies, they are reported in Table \ref{t:mpc_param}, in addition to the following settings: (i) prediction horizon $T=15$; (ii) number of samples $N_{tot}=95,472$; (iii) $\epsilon=0.95$; and (iv) $\delta=10^{-3}$. Moreover, the robustly stabilizing feedback gain matrix $K$ has been evaluated offline exploiting typical robust tools. The results presented in the following have been obtained exploiting a simulator developed with MATLAB/Simulink 2016b run in quasi real-time (i.e. $1$ s of simulation in about $1$ s) over a Intel Core i$7-7500$U CPU $@2.70$ GHz with $16$ GB of RAM and $512$ GB solid-state drive.

\begin{table}[!h]
\renewcommand{\arraystretch}{1.3}
\caption{SMPC weight matrices.}
\label{t:mpc_param}
\centering
\begin{tabular}{c c}
\hline \hline
Parameter & Value\\
\hline
$diag(Q_{long})$ & $[10^6,4\times 10^1,4\times 10^1,10^5,10^5]$\\
$diag(R_{long})$ & $[4\times 10^2, 3\times 10^{-3}]$\\
$diag(Q_{lat})$ & $[10^1,10^1,10^1,10^6]$\\
$R_{lat}$ & $10^6$\\
\hline \hline
\end{tabular}
\end{table}

In terms of uncertainties included in the model, a $\pm30\%$ variation of the following parameters has been considered, in addition to those parametric uncertainties ascribable to neglected nonlinearities: (i) cruise speed, with respect to the guidance and control flexibility to different flight conditions; (ii) FW-UAV mass, evaluating the possibility to adopt the same guidance and control strategy on slightly different vehicles; (iii) inertia, due to the manufacturing process. Moreover, we consider the presence of fixed-direction wind turbulence, representing a bounded persistent disturbance affecting the system and modeled as random with uniform distribution and maximum intensity of $\pm1$~m/s. The considered wind direction is reported in Figure \ref{fig:miami_sil}. 

\begin{table}[!h]
\renewcommand{\arraystretch}{1.3}
\caption{Additive disturbances considered.}
\label{t:additive_dist}
\centering
\begin{tabular}{c c | c c}
\hline \hline
\splitcell{ LONG.\\disturbance} & \splitcell{ Disturbance \\ value}  & \splitcell{ LAT.-DIR. \\disturbance} & \splitcell{ Disturbance \\ value} \\
\hline
$d_u$ [m/s] & $1.5$ & $d_v$ [m/s] & $1.5$\\
$d_{\alpha}$ [rad] & $10^{-2}$ & $d_p$ [rad/s] & $10^{-2}$\\
$d_{\theta}$ [rad] & $10^{-2}$ & $d_r$ [rad/s] & $10^{-2}$\\
$d_q$ [rad/s] & $10^{-2}$ & $d_{\phi}$ [rad] & $10^{-2}$\\
$d_h$ [m] & $10^{-1}$\\
\hline \hline
\end{tabular}
\end{table}

Additional external disturbances have been considered affecting the other states, with an analogous uniform distribution even if with a minor impact. The reader is referred to Table \ref{t:additive_dist} for further details.

\begin{table}[!h]
\renewcommand{\arraystretch}{1.3}
\caption{State constraint polytope vertexes.}
\label{t:vincoli}
\centering
\begin{tabular}{c c | c c}
\hline \hline
\splitcell{ LONG.\\variable} & \splitcell{Constraint \\ boundaries}  & \splitcell{ LAT.-DIR. \\variable} & \splitcell{Constraint \\ boundaries} \\
\hline
$\delta_u$ [m/s] & $\pm2$ & $\delta_v$ [m/s] & $\pm2$\\
$\delta_{\alpha}$ [rad] & $\pm10^{-1}$ & $\delta_p$ [rad/s] & $\pm10^{-1}$\\
$\delta_q$ [rad/s] & $\pm10^{-1}$ & $\delta_r$ [rad/s] & $\pm10^{-1}$\\
$\delta_{\theta}$ [rad] & $\pm10^{-1}$ & $\delta_{\phi}$ [rad] & $\pm10^{-1}$\\
$\delta_h$ [m] & $\pm2$\\
\hline \hline
\end{tabular}
\end{table}

\begin{figure}[htb]
\centering
\includegraphics[width=1\columnwidth]{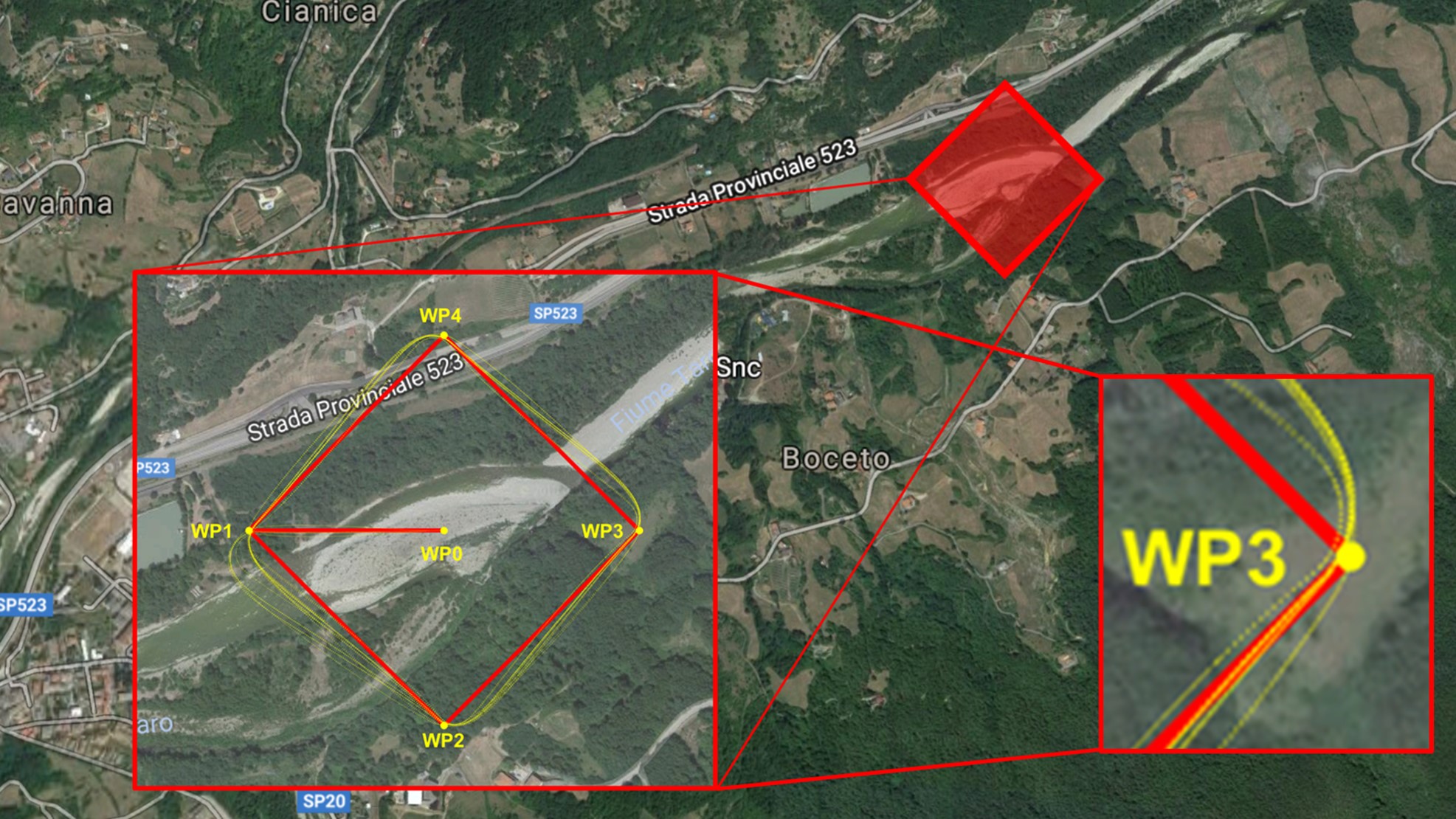}
\caption{20 different square-trajectories over the Taro River, Borgo Val di Taro, Parma, Italy. The area covered has an extension of about 180 ha.}
\label{fig:square_sil}
\end{figure}

\begin{figure}[htb]
\centering
\includegraphics[width=1\columnwidth]{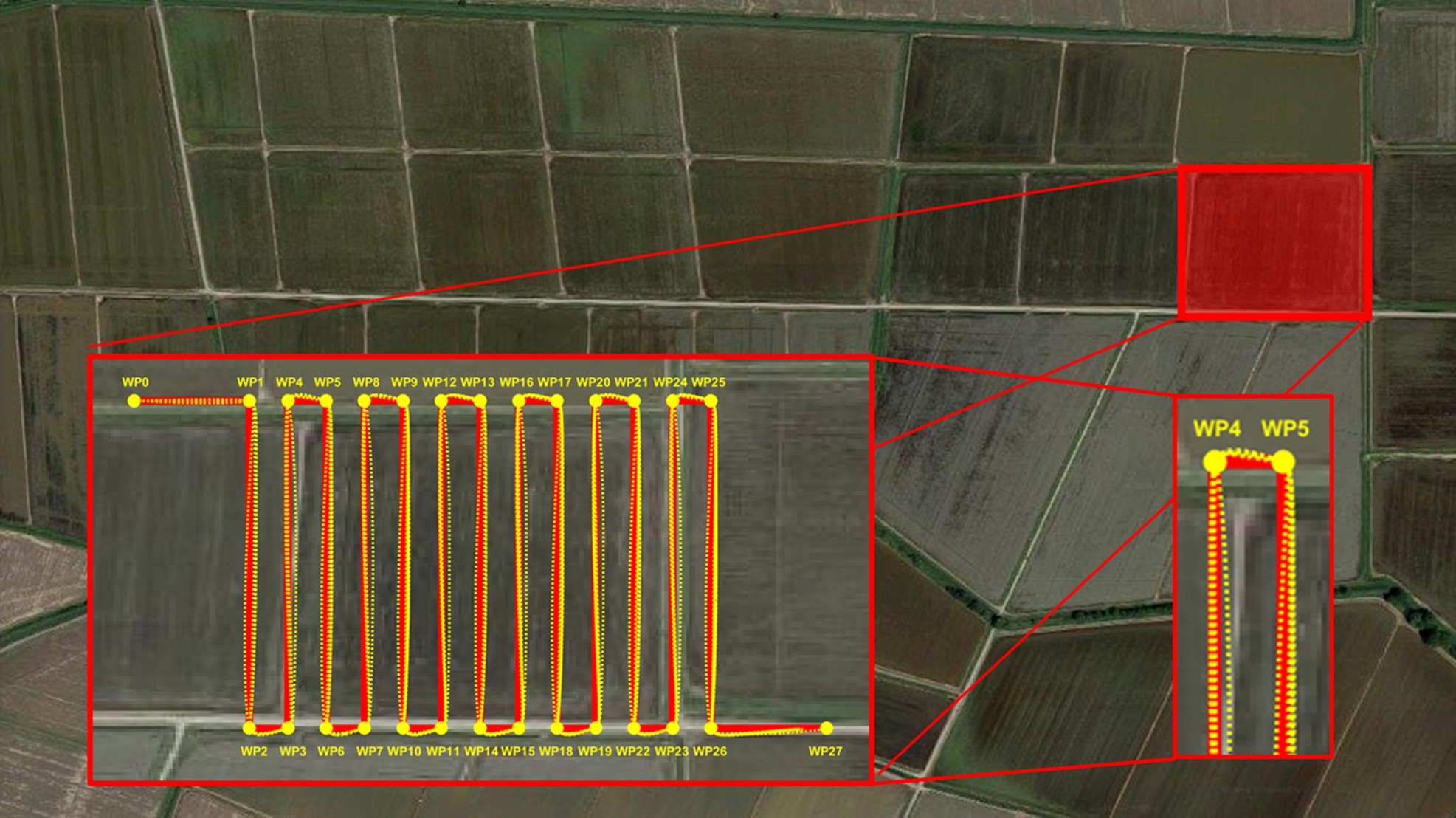}
\caption{20 different snake-trajectories over a paddy field in Olcenengo, Vercelli, Italy. }
\label{fig:snake_sil}
\end{figure}

Considering a time-varying reference $r_k$ provided by the guidance algorithm with respect to the waypoints setup for each scenario, the state deviations $\delta x_k$ have been constrained to be bounded within a polytope, which vertexes are reported in Table \ref{t:vincoli}. On the other hand, hard constraints have been imposed on the control variables in compliance with the vehicle mechanical limitations. Hence, the throttle $\Delta T$ can vary between $0\%$ and $100\%$, whereas the elevator and aileron deflection saturation was fixed to $\pm20\%$. The waypoints identification is strictly bounded to the mission scenario as well as the payload which the FW-UAV is equipped with. For each scenario, the path identified by the waypoints has been simulated 20 times and the results obtained are represented in Figure \ref{fig:square_sil}, Figure \ref{fig:snake_sil} and Figure \ref{fig:miami_sil}. Further details are provided in the following sections.
\subsection{Landslide Scenario}
In the frame of FW-UAV for civil protection, the selected area is a high-risk landslides area close to the Taro River, Parma, Italy $(45^{\circ}29'43.1''\text{N},9^{\circ}47'47.3''\text{E})$ (see Figure \ref{fig:caseA}), where the overfly is required to monitor the landslide movement in the Boceto area to prevent catastrophic consequences such as those occurred in February 2014. The 20 simulated square-trajectories over the critical area are represented in Figure \ref{fig:square_sil}.
\subsection{Paddy Field Scenario}
For precision farming, a paddy field at Olcenengo, Vercelli, Italy ($45^{\circ}22'22.2''\text{N}, 8^{\circ}17'34.3''\text{E}$) has been selected for exploiting FW-UAVs equipped with a OptRx$\textsuperscript\textregistered$ multi-spectral camera to obtain the NDVI of the rice field (see Figure \ref{fig:caseB}). A grid pattern allows to provide the required field coverage in compliance with the performance index of interest and the payload characteristics. Therefore, the grid width, identified by 27 waypoints, has been set in compliance with sidelap and overlap required by the active sensor itself, and including also an external band to allow the FW-UAV stabilization after each turn. Simulation results are depicted in Figure \ref{fig:snake_sil}.
The UAV flight mission is represented by a grid pattern, identified by 27 waypoints, over a $200$X$150$ m rectangular-shape area (see \ref{fig:caseA}). The grid-width is given by the camera performance with respect to the flight altitude. For $100$ m, the grid size has been set to $20$ m, including a $10\%$ of both overlap and sidelap requirements, following the sensor manufacture guidelines described in \cite{laura}. Moreover, the coverage area includes also a $10$ m band to allow the UAV stabilization after each turn.

\subsection{Urban Scenario}
A neighborhood in the Downtown Miami's Brickell Financial District, Miami, FL, USA $(25^{\circ}45'31.3''\text{N},80^{\circ}11'28.3''\text{W})$, around the Four Season Hotel and Tower, has been chosen for the urban scenario. Indeed, the Miami police enforcement is already evaluating the possibility to exploit FW-UAVs for drug interdiction, patrol missions, criminal surveillance as well as search-and-rescue operations, but only in life-threatening situations in imminent danger to life or properties. In this case, the selected waypoints have been identified with respect to the buildings-to-be-monitored location and the safety-area around them, set to be compliant with (possible) risk-mitigation rules (see Figure \ref{fig:caseC}). Considering a piecewise-linear pattern, the ideal path as well as the simulated ones are represented in Figure \ref{fig:miami_sil}.
\begin{figure}[htb]
\centering
\includegraphics[width=1\columnwidth]{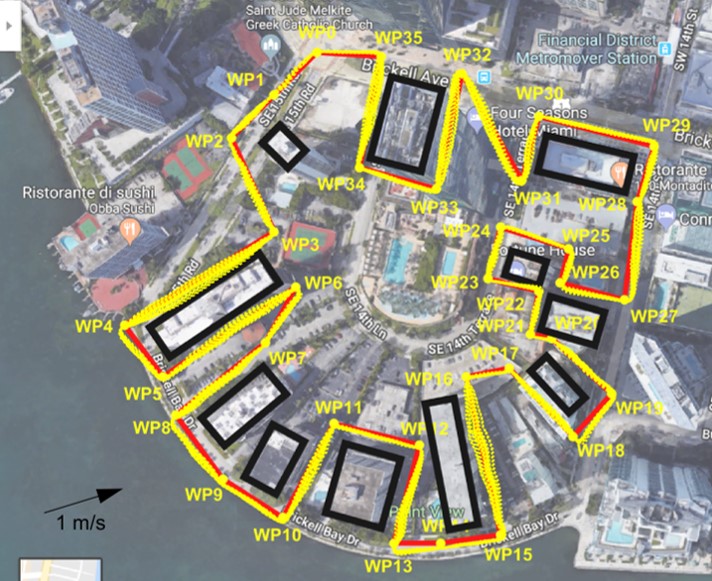}
\caption{20 different piecewise-linear trajectories over Downtown Miami's Brickell Financial District, around the Four Season Hotel and Tower - 2D trajectories.}
\label{fig:miami_sil}
\end{figure}
In this case, the selected waypoints have been identified with respect to the buildings-to-be-monitored location and the safety-area around them, set to be compliant with (possible) risk-mitigation rules. Considering a polygonal chain, which vertexes are identified by 36 waypoints, the ideal path (red line) as well as the simulated ones (yellow lines) are represented in Figure \ref{fig:miami_sil}, together with the wind turbulence considered, depicted by the black arrow at the bottom of the Figure.

\begin{figure}[h!]
\centering
\includegraphics[width=1\columnwidth]{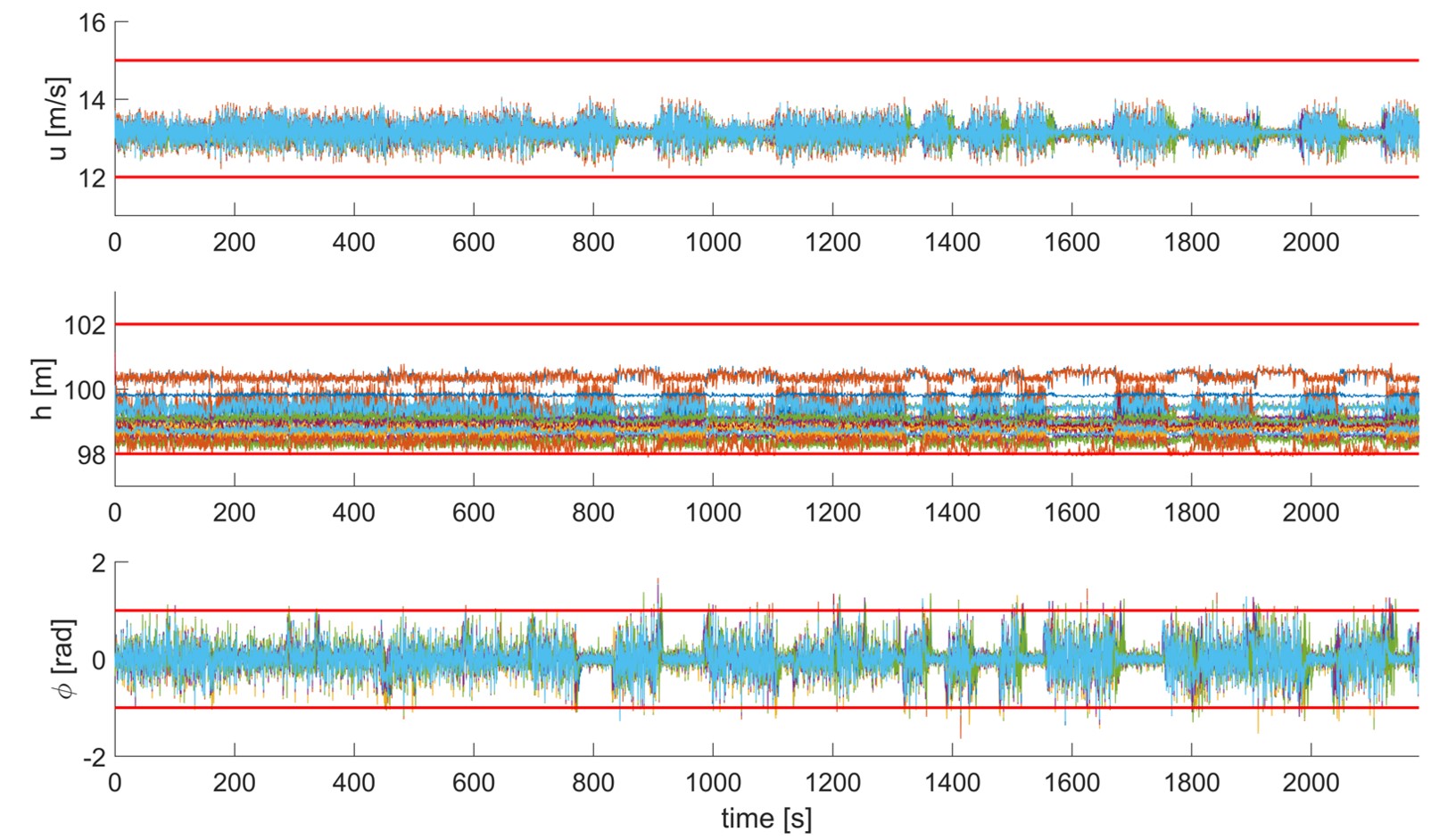}\label{fig:variables}
\caption{Airspeed, altitude and roll angle variation along 20 trajectories. The red lines represent the corresponding constraints imposed during the simulations.}
\end{figure}

In general, for all three scenarios, we can notice a "tangible" effectiveness of the proposed control strategy since the FW-UAV shows remarkable tracking capabilities with respect to the trajectory to follow. This is confirmed by Figure \ref{fig:variables} where the tracking capabilities are represented with respect to three state variables for the urban scenario: longitudinal component of the velocity $u$, altitude $h$ and roll angle $\phi$, within the corresponding constraint boundaries (red bold lines). As we can notice, only few values of the roll angle exceed the constraints, which is still acceptable with respect to the probability of constraint violation considered in this study.

\begin{figure}[h!]
\centering
\includegraphics[trim=0cm 0cm 1cm 1cm, clip=true,width=1\columnwidth]{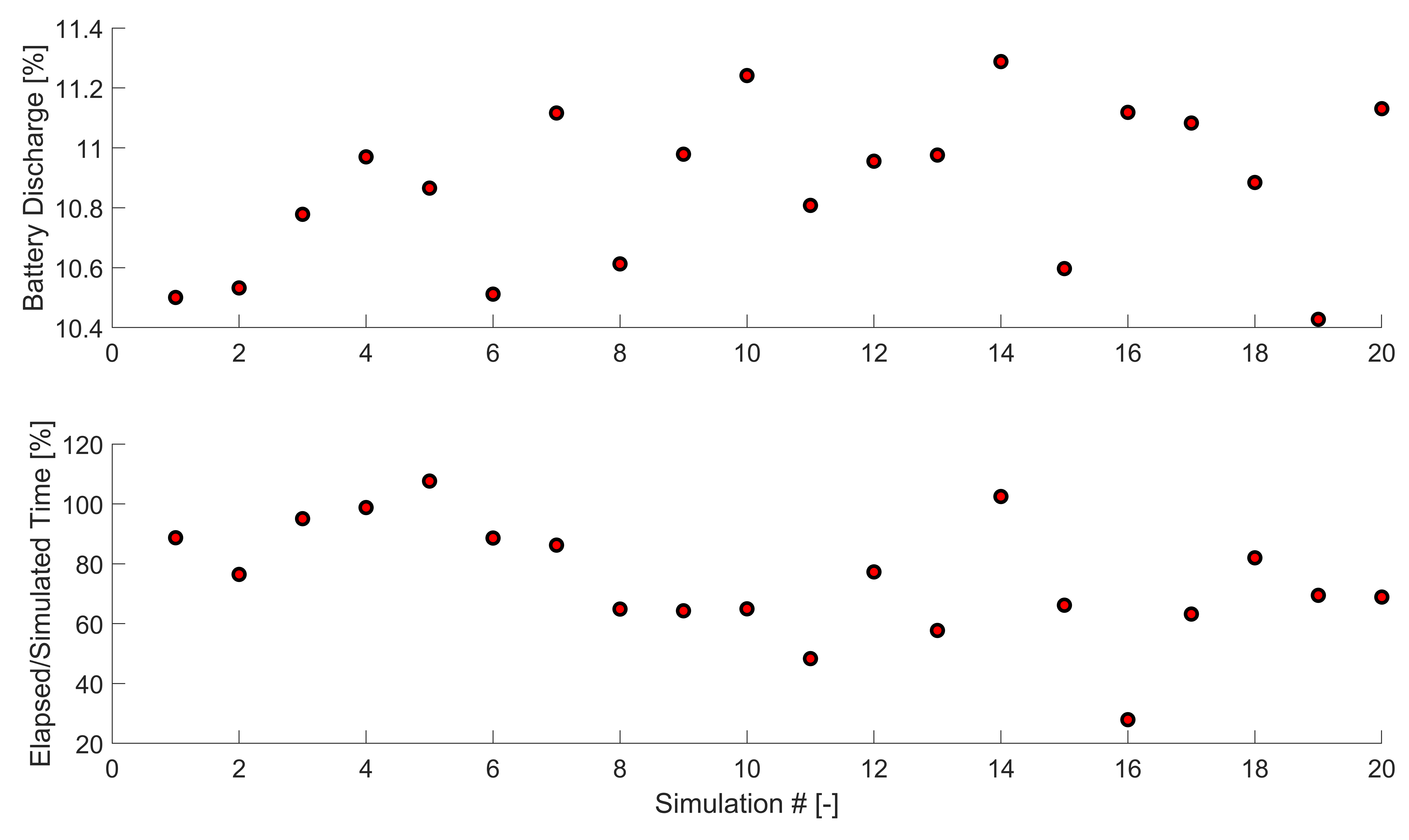}\label{fig:scatter}
\caption{Battery discharge percentage (upper) and computational effort (lower).}
\end{figure}

In order to provide a preliminary assessment of the SMPC control effort, only for the urban patrolling application, the battery discharge has been evaluated, and is represented in Figure \ref{fig:scatter} (upper graph) as a percentage of the total battery energy, i.e. 37 kJ, considering an initial electric charge of 5000 mAh and a nominal voltage of  7.4 V.
The actual current and voltage required by the system to the batteries have been estimated as a function of the torque provided by the propeller, following the Glauert's blade element theory, with respect to the throttle input control.

Moreover, the computational cost has been estimated for each simulation, by comparing the elapsed time (duration of the Simulink test) with respect to the simulated time (2180 s). As we can notice from Figure \ref{fig:scatter} (lower scatter plot), the elapsed time is lower than the simulated in the majority of cases, thus implying real-time implementability of the proposed scheme. On the other hand, future hardware-in-the-loop tests shall be performed to validate the computational compatibility of the chosen hardware with the presented controller. 

\begin{figure}[h!]
\centering
\includegraphics[trim=1cm 1cm 1cm 3cm, clip=true,width=1\columnwidth]{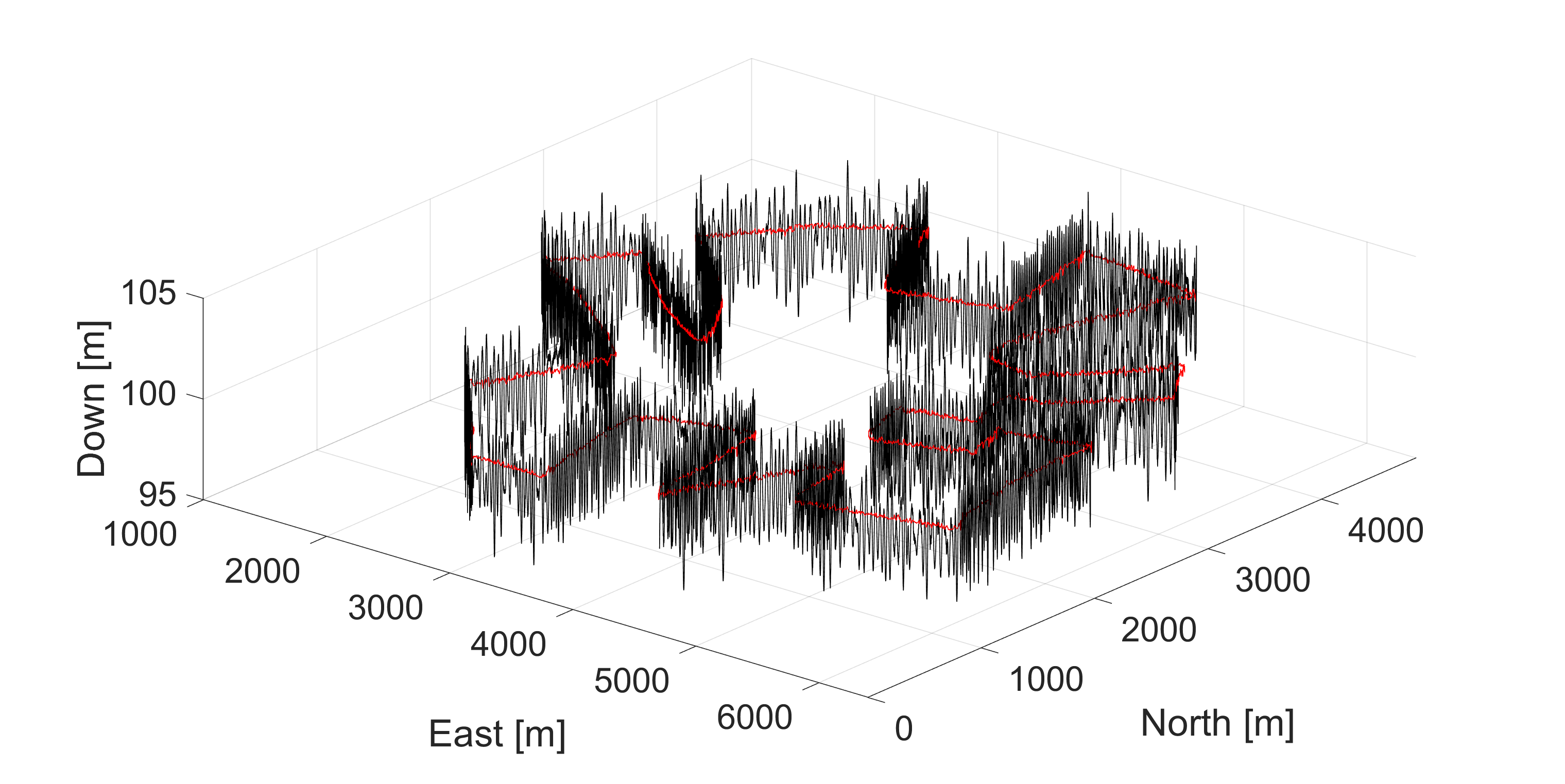}\label{fig:LQ_vs_S}
\caption{LQMPC 3D trajectory (black line) and SMPC 3D trajectory (red line).}
\end{figure}

To complete the SMPC scheme validation, the proposed controller effectiveness has been compared with a classical MPC, still considering the presence of a fixed-direction wind turbulence of 1 m/s. Figure \ref{fig:LQ_vs_S} represents both the trajectories obtained, i.e red for SMPC and black for LQMPC, considering the same MPC parameters in terms of Q and R matrices as well as prediction horizon. Apparently, both control approaches show an effective response in the presence of wind turbulence. However, to properly control the FW-UAV during the trajectory, it was necessary to reduce the LQMPC sample time to 0.01 s instead of 0.1 s. Hence, the computation cost raised to 140\%, i.e. the elapsed time is about 3057 s. Analogously, also the control effort results to be much higher, with a residual battery energy of about 46\%.

\section{Conclusions}
In this research, a guidance algorithm and a tracking SMPC based on offline sampling are combined to provide guidance and control capabilities in the presence of additive noise (i.e. wind turbulence) and uncertainties (i.e. model parameter variations). In particular, the approach has been validated considering three different mission scenarios: (i) monitoring of a landslide area ; (ii) precision farming of a paddy-field; and (iii) a patrolling mission over a Miami's Downtown urban area. The effectiveness of the strategy proposed is validated by software-in-the-loop tests, evaluating both the control effort in terms of battery discharge and computational cost for the most demanding missions, i.e. the urban monitoring. The results obtained in all three cases show the SMPC capability of guaranteeing good stability performance of the platform and remarkable tracking capabilities. Moreover, the preliminary results are promising for future hardware-in-the-loop validation. To conclude the analysis, the effectiveness of the SMPC approach has been proved through its comparison with a classical MPC scheme, still for the urban monitoring.

\addtolength{\textheight}{-12cm}   


\bibliographystyle{IEEEtran}
\bibliography{BIBLIO}

\end{document}